\documentclass[sigchi]{acmart}

\usepackage{rotating}
\usepackage{booktabs} 
\usepackage{arydshln}
\usepackage{pbox}
\usepackage{dsfont} 

\newcommand{\rev}[1]{{\color{black}#1}\normalfont}

\newcommand{\etal}{et al.}

\newcommand{\OurName}{$K$-modal GAN\xspace}
\newcommand{\SName}{sMM-GAN\xspace}
\newcommand{\UName}{uMM-GAN\xspace}

\setcitestyle{square}

\usepackage{xcolor,colortbl}
\usepackage{multirow}
\usepackage{graphicx}
\usepackage{wrapfig}

\usepackage{caption}
\usepackage{subfig}

\usepackage[ruled]{algorithm2e} 

\SetAlFnt{\small}
\SetAlCapFnt{\small}
\SetAlCapNameFnt{\small}
\SetAlCapHSkip{0pt}
\usepackage[noend]{algpseudocode}
\usepackage{enumitem}
\acmJournal{TOG}

\begin{document}
\title{Unsupervised Multi-modal Styled Content Generation}

\begin{abstract}
The emergence of deep generative models has recently enabled the automatic generation of massive amounts of graphical content, both in 2D and in 3D.
Generative Adversarial Networks (GANs) and style control mechanisms, such as Adaptive Instance Normalization (AdaIN), have proved particularly effective in this context, culminating in the state-of-the-art StyleGAN architecture.
While such models are able to learn diverse distributions, provided a sufficiently large training set, they are not well-suited for scenarios where the distribution of the training data exhibits a multi-modal behavior.
In such cases, reshaping a uniform or normal distribution over the latent space into a complex multi-modal distribution in the data domain is challenging, and the generator might fail to sample the target distribution well.
Furthermore, existing unsupervised generative models are not able to control the mode of the generated samples independently of the other visual attributes, despite the fact that they are typically disentangled in the training data.  

In this paper, we introduce \UName, a novel architecture designed to better model multi-modal distributions, in an unsupervised fashion. 
Building upon the StyleGAN architecture, our network learns multiple modes, in a \emph{completely unsupervised manner}, and combines them using a set of learned weights.
We demonstrate that this approach is capable of effectively approximating a complex distribution as a superposition of multiple simple ones.
We further show that \UName effectively disentangles between modes and style, thereby providing an independent degree of control over the generated content.
\end{abstract}

\begin{teaserfigure}
\centering
    \includegraphics[scale=0.775]{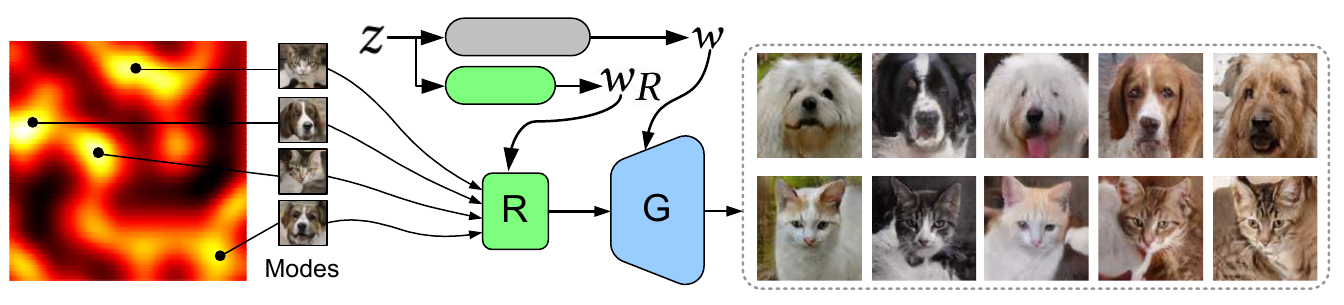}
    \caption{Our unsupervised $K$-modal GAN (\UName with $K=4$) trained on a dataset of images of cats and dogs, with various visual attributes shared by both kinds of images.
   	The complex distribution of the training data is approximated by the generator by sampling a mixture of $K=4$ modes, which are learned in an unsupervised manner. In this case, two of the modes (visualized on the left) correspond to cats, and two to dogs. Thus, at test time it is possible to control the species in the generated samples. Furthermore, the mode and the style in our generator are disentangled, enabling changing one while preserving the other, as demonstrated by the generated samples on the right. In each column the mode is switched, while keeping the style parameters fixed, yielding similar fur colors and patterns. 
	}
  \label{fig:teaser}
\end{teaserfigure}


\author{Omry Sendik}
\orcid{0000-0002-9268-8281}
\affiliation{%
 \institution{Tel Aviv University}
 \country{Israel}}
\email{omrysendik@gmail.com}
\author{Dani Lischinski}
\affiliation{%
 \institution{Hebrew University, Jerusalem}
  \country{Israel}
}
\author{Daniel Cohen-Or}
\affiliation{%
 \institution{Tel Aviv University}
 \country{Israel}}

\maketitle


%
%


\section{Introduction}

Content generation has been a major bottleneck since the dawn of computer graphics. 
Recently, the emergence of generative models based on deep neural networks, finally carries a promise for being able to automatically generate massive amounts of diverse content.
Although the visual quality of deep generative models could not initially rise up to the high visual fidelity bar of the field, it has been improving rapidly.
Some of the most promising approaches, in terms of visual fidelity are Generative Adversarial Networks (GANs) \cite{goodfellow2014generative}, which learn to generate samples whose distribution closely resembles that of the training data.

While GANs are able to generate a large amount of varied data, provided a sufficiently large training set, they are not explicitly designed for scenarios where the distribution of training data exhibits a multi-modal behavior. Consider, for example, a dataset consisting of several different species of animals, or several different kinds of cars. In such cases, reshaping a simple distribution over the latent space into a complex multi-modal one is challenging, and the learned distribution might fail to approximate that of the training data. A number of works note and attempt to address this issue, as briefly reviewed in Section~\ref{sec:background}.

Furthermore, the control that GANs typically provide over the generated data is limited, especially in the unsupervised learning scenario, i.e., when the data comes without any additional annotations. A degree of control may be achieved by adding conditioning \cite{mirza2014conditional}. However, this requires supervision, which is not always feasible.

Karras \etal~\shortcite{karras2019style} recently introduced StyleGAN, an unsupervised framework that leverages the AdaIN style control mechanism \cite{huang2017arbitrary} to afford some degree of control over attributes at different scales. By first mapping a simply distributed latent space into an intermediate one, they are able to better approximate the training probability density, while avoiding entangling the factors of variation. The StyleGAN architecture has resulted in the highest quality data-driven generative models to date.
Note, however, that continuous style parameters are not well suited for transitions between distinct modes in the distribution.
Although the different modes are typically disentangled from the other visual attributes in the data, neither StyleGAN, nor any other unsupervised generative model, provides means for controlling the mode of the generated samples independently of the other attributes.

In this work, we introduce a new generative model, designed explicitly for coping with multi-modal distributions.
For such distributions, our model results in a better approximation of the training distribution, in addition to providing the ability to directly control the mode of the generated samples, independently of the control afforded by the style parameters. 
Remarkably, our approach requires no supervision; provided only the number of modes $K$ as a hyper-parameter, it is able to learn the different latent modes in the training distribution, while learning to generate samples resulting from them.

Specifically, the architecture that we introduce here is based on that of StyleGAN \cite{karras2019style}. The StyleGAN generator uses a single learned constant that serves as a common seed or \emph{root} for all of the synthesized samples from a given distribution. All the variations among the generated samples result from varying AdaIN parameters, as well as random noise inputs, at each generation scale.
In contrast, we propose to learn \emph{$K$ root constants}, which may be interpreted as seeds for the different modes present in the data. Intuitively, rather than relying on the ability of the mapping network to deform the latent space distribution into one that follows the training probability density, we approximate the latter using a composition of simpler distributions around multiple modes.

We show that our $K$-rooted generators are able to better approximate multi-modal distributions, compared with the state-of-the-art.
Perhaps more importantly, we show that the root constants in our architecture correspond to different modes in the training set.
Finally, spanning the learned distribution in this manner,
results in an explicit method for controlling the mode of the generated data, providing a new axis of control which is natural for multi-modal distributions, and has little or no effect on the attributes controlled via AdaIN and noise, as demonstrated in Figure~\ref{fig:teaser}. Thus, our approach effectively disentangles mode and ``style''.

\section{Background}
\label{sec:background}

\begin{figure}[t]
	\centering
	\begin{tabular}{c}
		\includegraphics[scale=0.5]{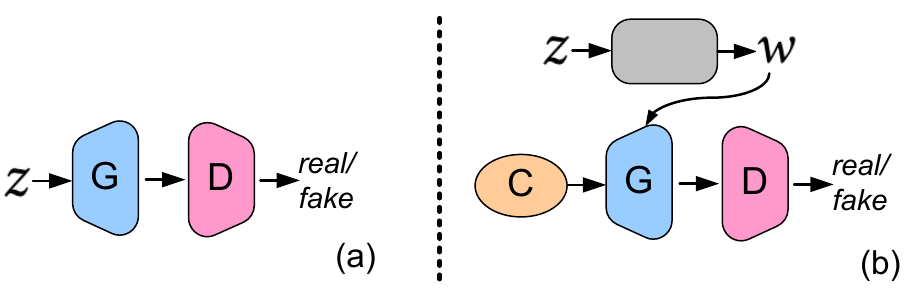}
	\end{tabular}
	\caption{(a) The original GAN architecture proposed by Goodfellow \etal~\shortcite{goodfellow2014generative}. (b) The StyleGAN of Karras \etal~\shortcite{karras2019style}, which is composed of a mapping network (from $z \in \mathcal{Z}$ to $w \in \mathcal{W}$) and a synthesis network, fed by a learned root constant $C$. The synthesis is controlled using AdaIN parameters derived from $w$.
	}
	\label{fig:Archs}
\end{figure}

We explore the problem of unsupervised generation of natural images.
Research in this area received a tremendous boost with the introduction of Generative Adversarial Networks (GANs) \cite{goodfellow2014generative}.
Given a set of training samples, GANs learn to generate samples whose distribution closely resembles that of the training data.
The commonly used DCGAN architecture \cite{radford2015dcgan}, outlined in Figure \ref{fig:Archs}(a), consists of a convolutional generator $G$, fed by a random input vector $z \in \mathcal{Z}$, typically drawn from a normal or a uniform distribution.
The generator is adversarially trained using a convolutional discriminator $D$.

Such GAN architectures are able to learn to sample sufficiently restricted distributions, however, they are not well suited for distributions that exhibit significant diversity or multi-modal behavior.
Multi-modal distributions may be sampled using conditional GANs \cite{mirza2014conditional}, provided that the training data is suitably annotated, for example with class labels.
For example, Odena \etal~\shortcite{odena2016conditional} concatenate a one-hot class vector to the generator's noise input.
Brock \etal~\shortcite{brock2019biggan} have demonstrated that class-conditioning makes it possible to effectively sample distributions as diverse as ImageNet \cite{ILSVRC15}.

This limitation of GANs has been discussed by multiple researchers (e.g., \cite{gurumurthy2017deligan,ben2018gaussian,khayatkhoei2018disconnected,pandeva2019mmgan, xiao2018bourgan}).
Several works, such as \cite{ben2018gaussian,gurumurthy2017deligan,pandeva2019mmgan}, cope with the multi-modal case 
by structuring the latent space $\mathcal{Z}$ as a Mixture of Gaussians, whose first and second moments are learned.
\rev{Xiao et al.~\cite{xiao2018bourgan} also use a Mixture of Gaussians in order to avoid mode collapse.} 
While some of these approaches are unsupervised, they impose a specific parametric model on the latent distribution, and mostly assume that the data consists of disconnected clusters or manifolds. Contrary to these works, our approach makes no a priori assumptions regarding the form of the distribution (i.e., we do not assume it is a Gaussian Mixture Model). Neither do we assume that the distribution consists of several disconnected parts.
In addition, our approach is unique in that it enables disentangled control of mode and other visual attributes.

Brock \etal~\cite{brock2019biggan} note that the choice of latent space might significantly affect the performance of GANs. In particular, they observed that the discrete Bernoulli $\{0,1\}$ latent space outperformed the normal distribution (without truncation), arguing that it might reflect the prior that the underlying factors of variation in natural images are discrete (one feature is present, another is not). In our approach there is no need to choose between continuous and discrete distributions, since both the modes and their mixture are learned by the network. Thus, the network adapts itself to the training data, rather than being pre-designed for a particular kind of distribution.

In addition to GANs, a variety of alternative generative models have also been explored. 
Notable examples include Variational Auto-Encoders (VAEs) \cite{kingma2014vae}, and variants thereof that attempt to achieve clustering in the latent space, such as GMVAE \cite{dilo2016gmmvae}, or ClusterGAN \cite{mukherjee2019clustergan}, which combines the advantages of VAEs and GANs.
Another unique work in the line of Generative models is Glow \cite{kingma2018glow}. By using 1x1 invertible convolutions within a flow-based generative model, Kingma and Dhariwal have demonstrated that a generative model is capable of efficiently synthesizing realistic-looking images.
None of the above methods, however, were able to demonstrate the high-resolution sharp images, such as those that may be produced by the latest GANs.

\rev{It should be noted that the multi-modal case has also been addressed in the context of Image-to-Image translation~\cite{zhu2017toward,huang2018multimodal}, which is a rather different setting where the generation is conditional, and the training requires paired data \cite{zhu2017toward} or separate datasets \cite{huang2018multimodal}.}

\subsection{StyleGAN}

The state-of-the-art in GAN-based image generation is achieved by the StyleGAN architecture \cite{karras2019style}, a fascinating combination between GANs and the AdaIN mechanism \cite{huang2017arbitrary}, originally introduced in the context of style transfer.
The StyleGAN architecture, outlined in Figure \ref{fig:Archs}(b), first maps the latent vectors $z$, normally or uniformly distributed in a latent space $\mathcal{Z}$, into a properly shaped intermediate latent representation $w \in \mathcal{W}$, using a learned mapping network.
The $w$ vectors are then transformed into sets of AdaIN parameters, which are injected into the different levels of the generator. 

\begin{figure}[t]
	\centering
	\begin{tabular}{cccc}
	    \includegraphics[height=0.19\columnwidth]{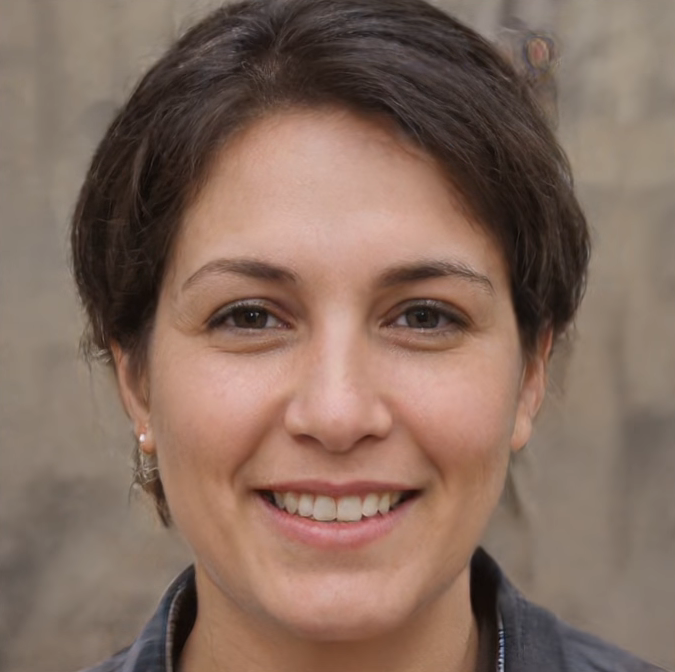} & 
	    \includegraphics[height=0.19\columnwidth]{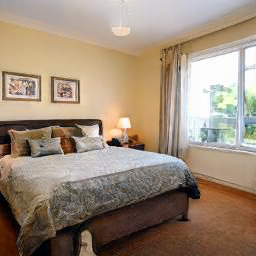} &
	    \includegraphics[height=0.19\columnwidth]{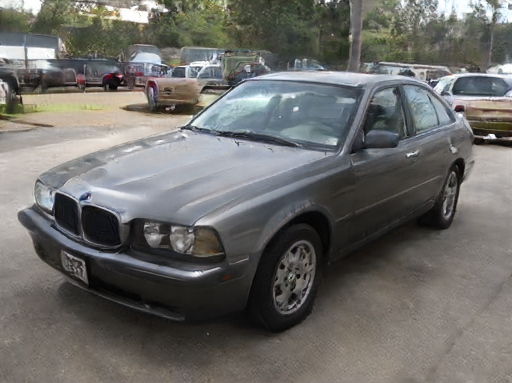} &
	    \includegraphics[height=0.19\columnwidth]{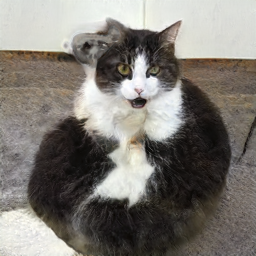}
	\end{tabular}
	\caption{Images generated by different StyleGAN models using an average style vector $w_\textrm{avg}$ (obtained by drawing a large number of $z$ vectors and averaging their corresponding $w$ vectors).
	}
	\label{fig:StyleGAN-constants}
\end{figure}

Notably, the generator takes as input a single (learned) constant, which we refer to as the \emph{root constant}. The root constant may be seen as an encoding of the learned mode of the training data. Figure~\ref{fig:StyleGAN-constants} visualizes this learned mode for several StyleGAN models trained by Karras \etal~on different datasets (faces, bedrooms, cars, and cats). The visualization is obtained by feeding the generator network $G$ with an average style vector $w_{\textrm{avg}}$, and without noise inputs. For example, note that the generated face shown in Figure~\ref{fig:StyleGAN-constants} is neither very masculine nor very feminine and its complexion is not very dark, nor very pale. The generator is only able to shift away from these average properties through the use of different sets of AdaIN parameters, obtained by transforming different vectors $w \in \mathcal{W}$. Thus, to model a given distribution well, the shape of the distribution should be echoed by that of the intermediate latent space $\mathcal{W}$.

\begin{figure}
	\includegraphics[scale=0.121]{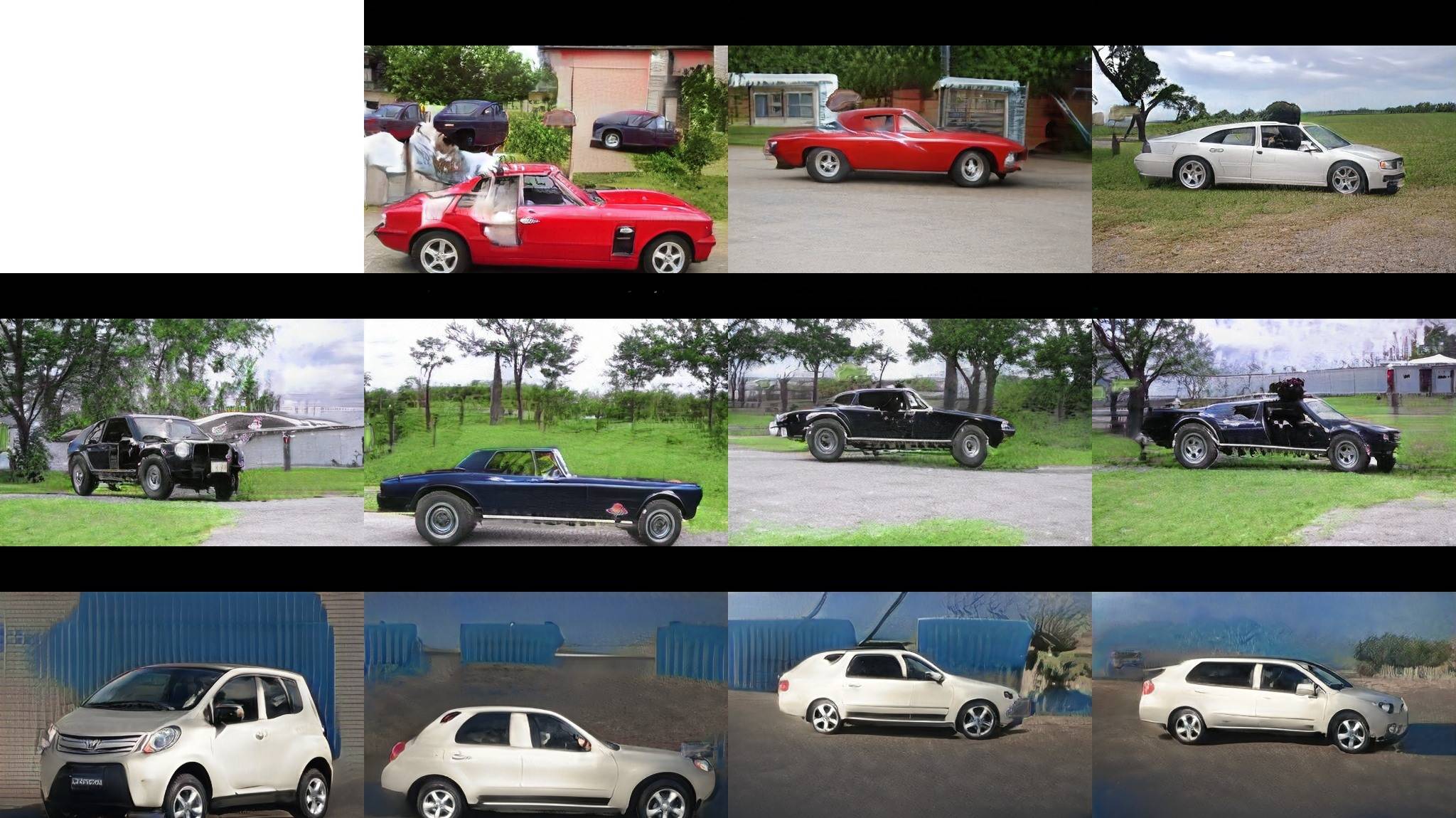}
	\caption{\label{fig:StyleGANPoseMix}
		Style mixing in StyleGAN. The coarse level style parameters are taken from the images in the top row, while the style parameters for the other generator layers are taken from the images in left column.
		The results of such mixing are inconsistent and unpredictable. While the car poses of the mixtures indeed come from the top row, one car in the 2nd row is blue, while the other two are black, and their shapes seem strongly affected by those of the cars in the top row. In contrast, in the 3rd row the shape is more strongly influenced by the one on the left, but it changes significantly between the different poses.
	}
\end{figure}

StyleGAN has been shown to perform admirably on several image domains, however, we note that it is still best suited for distributions with a single dominant mode, as we shall demonstrate in the next section.
In contrast, in this work we are interested in multi-modal domains, where the modes might correspond to distinct peaks in the probability density, or might span the probability density in a more continuous manner.
In either case, the training data samples typically share some common attributes, which span across modes.
Consider a set of fonts, for example. Within each font, letters share common properties, such as font-weight, style and character dimensions. However, the distribution is multi-modal, where each character defines a separate mode.
In cases such as this, the distribution is unlikely to be well approximated by a set of multiscale variations around a single mode.
Consequently, attempting to train a single model on the entirety of such a dataset (without supervision such as class-conditioning) might lead to implausible results, as demonstrated in Figure \ref{fig:StyleGANvsMultiConstGAN}(a).

Furthermore, although mapping the latent space $\mathcal{Z}$ to an intermediate latent space $\mathcal{W}$, before obtaining the style parameters for the different generator levels, achieves some degree of disentanglement, this is not enough to provide a consistent and predictable way of controlling a fixed set of high-level attributes, such as pose or color. This is demonstrated in Figure \ref{fig:StyleGANPoseMix}, where we attempt to use the style mixing mechanism, proposed by Karras \etal~\shortcite{karras2019style} to control the pose of the cars in the left column using the poses of the cars in the top row. While the pose is indeed controlled, the results show that it is entangled with other attributes. In this work, we achieve a consistent disentanglement between the aspect captured by the different modes and the other visual attributes.

Several recent works demonstrated how a pretrained generative model, such as StyleGAN, may be used for image editing, by extracting the latent space embedding of an image that one wishes to manipulate~\cite{abdal2019image2stylegan,gabbay2019style}. 
Given this embedding, one may apply linear operations between latent vectors in order to manipulate the output result. A drawback of such approaches is that unsupervised GANs often make it difficult to control the desired attributes. By disentangling the embedding vector via an affine mixture assumption, Gabbay and Hoshen \shortcite{gabbay2019style} achieve control over face properties. Their approach relies on the observation that the average pose of a face is frontal. These recent approaches are related to our paper, since they also aim to achieve unsupervised disentanglement and thus control over the final result. Our approach however, is fundamentally different in the sense that we train a new unsupervised model, and achieve disentanglement through multi-modal learning. It is neither limited to faces, nor does it make any data dependent assumptions.

\section{Mode Mixture Modeling}
\label{sec:method}



As discussed earlier, StyleGAN samples the learned distribution by injecting style parameters and noise, at multiple scales, always starting from a single root constant, which encodes a single mode of the distribution.
Instead, we aim to represent the learned distribution using a model which is an explicit mixture of modes.
Rather than expecting the mapping network to learn a complex deformation of the latent space, our architecture explicitly reflects the fact that the distribution is modeled using $K$ modes, by training a generator that takes $K$ learned root constants as input, instead of a single one. The $K$ constants are mixed using a set of learned weights and the resulting mixtures are further modified by an independent set of style (AdaIN) parameters and noise inputs, to generate samples that cover the entire learned distribution. If the distribution is roughly discrete, and the training data is annotated accordingly, the corresponding constants and their mixture may be learned in a supervised fashion. However, as we show in Section~\ref{sec:unsupervised}, supervision is actually not necessary. 

\begin{figure}[t]
    \centering
    \begin{tabular}{c}
        \includegraphics[scale=0.80]{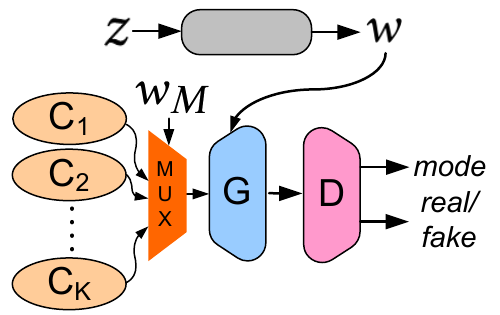} \\
        \hline\\
        \includegraphics[scale=0.80]{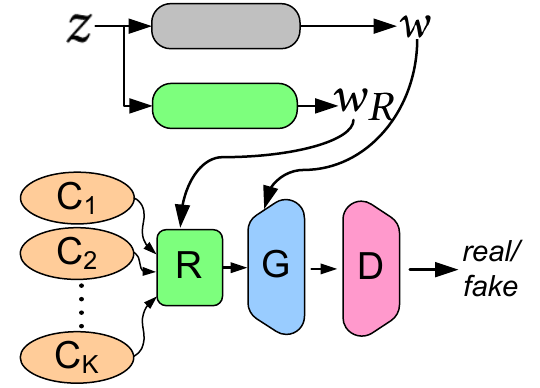}
    \end{tabular}
    \caption{Our \OurName architecture borrows from StyleGAN the generator $G$, discriminator $D$ and the mapping network that maps $z \in \mathcal{Z}$ to $w \in \mathcal{W}$. In place of a single constant input to $G$, $K$ root constants are learned. \textbf{Top}: In the supervised \SName, one of the $K$ constants is selected by a multiplexer and forwarded to $G$. Note that the discriminator penalizes the generator at train time on its generated modes.
    \textbf{Bottom:} In the unsupervised \UName, the $K$ learned constants are combined by $R$, using a vector of weights $w_R$ to yield the input of $G$. The mixture weights $w_R$ are generated by a mapping network, with a softmax as its last layer.} 
    \label{fig:arch}
    \label{fig:archMultiConst}
\end{figure}



\begin{figure}[t]%
	\centering
	\begin{tabular}{ccc}
		training data & StyleGAN & \UName $K = 3$ \\
		distribution & IoU = 0.53 & IoU = 0.86 \\	\includegraphics[scale=0.2]{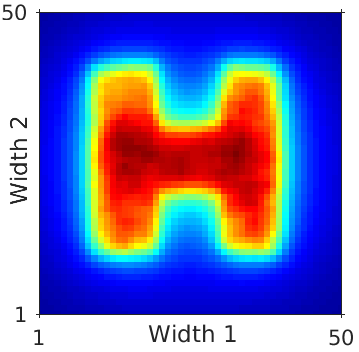} & \includegraphics[scale=0.2]{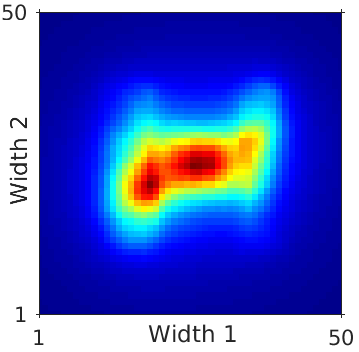} &
		\includegraphics[scale=0.2]{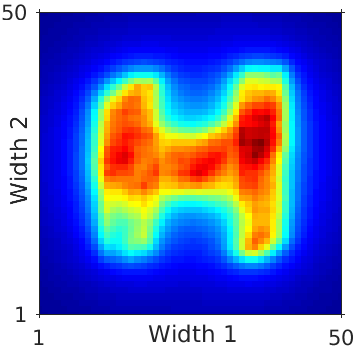} \\ \hline \vspace{-2mm} \\
		\includegraphics[scale=0.2]{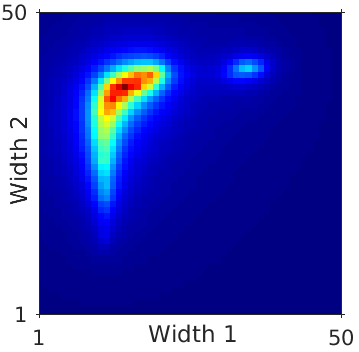} &
		\includegraphics[scale=0.2]{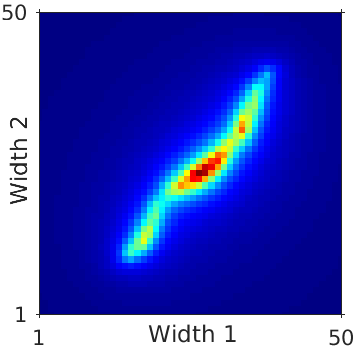} &
		\includegraphics[scale=0.2]{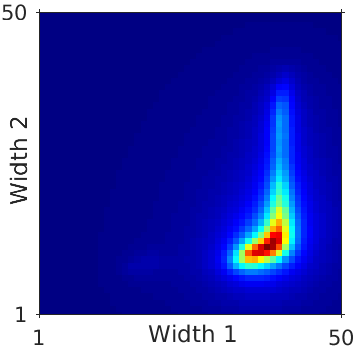} \\ 
		\hspace{1mm}
		\includegraphics[scale=0.94]{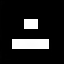} &
		\hspace{1mm}
		\includegraphics[scale=0.94]{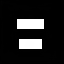} &
		\hspace{1mm}
		\includegraphics[scale=0.94]{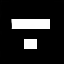} \\
		Mode 1 & Mode 2 & Mode 3 \\
	\end{tabular}
	\caption{Effect of multi-modal generation. \rev{We form a dataset where each data sample is an image with two white rectangles of the same height, but varying widths, such as the images shown in the bottom row. The widths provide a 2D parameterization, and the distribution of the training samples forms an H-shape in the plane (top left).} Training StyleGAN on this dataset reproduces only a part of the distribution (top middle), while training \UName with $K=3$ results in a much better approximation, and a higher Jaccard index value (IoU). The three learned modes are visualized in the middle and bottom rows. The middle row visualizes the distributions that are sampled by replacing the mixture weights vector $w_R$ with a one-hot vector. The combination of the resulting distributions is able to reproduce the H-shape. The bottom row shows the generated samples corresponding to each of the three learned constants.}
	\vspace{-1em}
	\label{fig:RectsH_PDF_Modes}
\end{figure}

\subsection{\OurName}
\label{sec:unsupervised}

We begin by describing \SName, a supervised version of our $K$-modal GAN, which may be suitable for the case where the data modes correspond to distinct clusters, and each training sample is annotated with the mode to which it belongs. This supervised architecture is depicted in the top diagram of Figure~\ref{fig:archMultiConst}. The architecture borrows from StyleGAN its generator $G$ and the mapping network that maps $\mathcal{Z}$ to $\mathcal{W}$. However, in place of a single constant input to $G$, multiple root constants, denoted $C_1, \ldots, C_K$ are learned, which are intended to encode up to $K$ modes of the distribution. A one-hot vector $w_M$ controls a simple multiplexer that chooses one of these root constants to be forwarded as input to $G$.
In order to learn $K$ constants that would correspond to $K$ modes in the data, we add a $K$-way classifier to the discriminator $D$, which is used to penalize the generator when it fails to produce an image that does not belong to the mode indicated by $w_M$.

Our main goal, however, is to learn the mode mixture model in an unsupervised fashion. This is accomplished by \UName, our unsupervised $K$-modal GAN, depicted in the bottom diagram of Figure~\ref{fig:arch}. As before, we learn $K$ root constants; however, in this variant, these constants are fed into a mixing layer (denoted $R$), where they are combined using $K$ weights $w_R$, and the resulting vector is fed into the generator $G$. The mixture weights $w_R$ are generated from the latent vectors $z \in \mathcal{Z}$ using a mapping network parallel to the one that maps $\mathcal{Z}$ to $\mathcal{W}$, and ended by a softmax layer that normalizes the mixture weights to a sum of one.

\rev{We demonstrate the advantage of our \UName over the original StyleGAN  using a simple synthetic example, where each data sample is an image with two white rectangles of the same height, but varying widths (such as the images shown in the bottom row of Figure \ref{fig:RectsH_PDF_Modes}). Thus, the widths of the two rectangles provide a natural embedding of the training set on the plane. The width of one rectangle (Width 1) is uniformly sampled from $[0,40]$, while the range of widths of the second rectangle (Width 2) depends on Width 1. The joint PDF of the two widths forms the shape of the letter H, as shown in the top left image of Figure \ref{fig:RectsH_PDF_Modes}.
Note that although this joint PDF is continuous, it nevertheless exhibits a complex anisotropic shape in the plane, which may be decomposed into multiple modes.}

Having drawn 10,000 samples from this distribution, the resulting set of images is used to train a StyleGAN, as well as a \UName (with $K=3$). Visualizing the empirical PDFs sampled by these two generative models, we can see that StyleGAN is unable to faithfully reproduce the training distribution (top row, middle). In contrast, the PDF of \UName with $K=3$ approximates the training PDF quite well (top row, right). In order to quantify the fit, we apply a threshold over the PDFs and compute the Jaccard index (Intersection over Union) between the training PDF and that of the two generated ones, resulting in an index of 0.86 for \UName, compared to 0.53 for StyleGAN. 

It is also instructive to examine the PDFs generated by \UName around each of the three modes that it has learned, visualized in the middle row of Figure \ref{fig:RectsH_PDF_Modes}. Note that each of these PDFs covers a different part of the H shape, facilitating the approximation of the anisotropic training PDF as a superposition of three simpler ones. Also note, that the shapes of the three PDFs are learned by the network, and they differ from a 2D Gaussian, as well as from each other.
The images corresponding to each of the three modes are shown in the bottom row of Figure \ref{fig:RectsH_PDF_Modes}.

\begin{center}
\begin{figure*}
    \begin{tabular}{cc} 
    $K=2$ &
    \begin{tabular}{c}
    \includegraphics[width=0.90\textwidth]{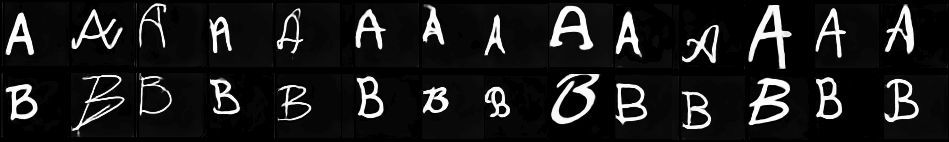}
    \end{tabular} \\
    $K=2$ &
    \begin{tabular}{c}
    \includegraphics[width=0.90\textwidth]{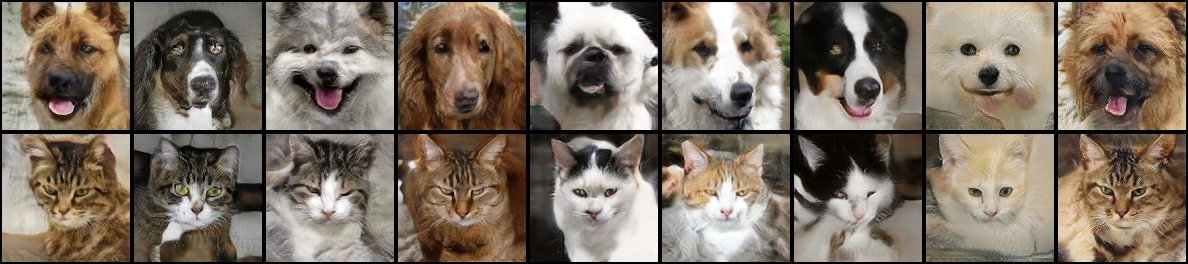}
    \end{tabular} \\
    $K=4$ &
    \begin{tabular}{c}
    \includegraphics[width=0.90\textwidth]{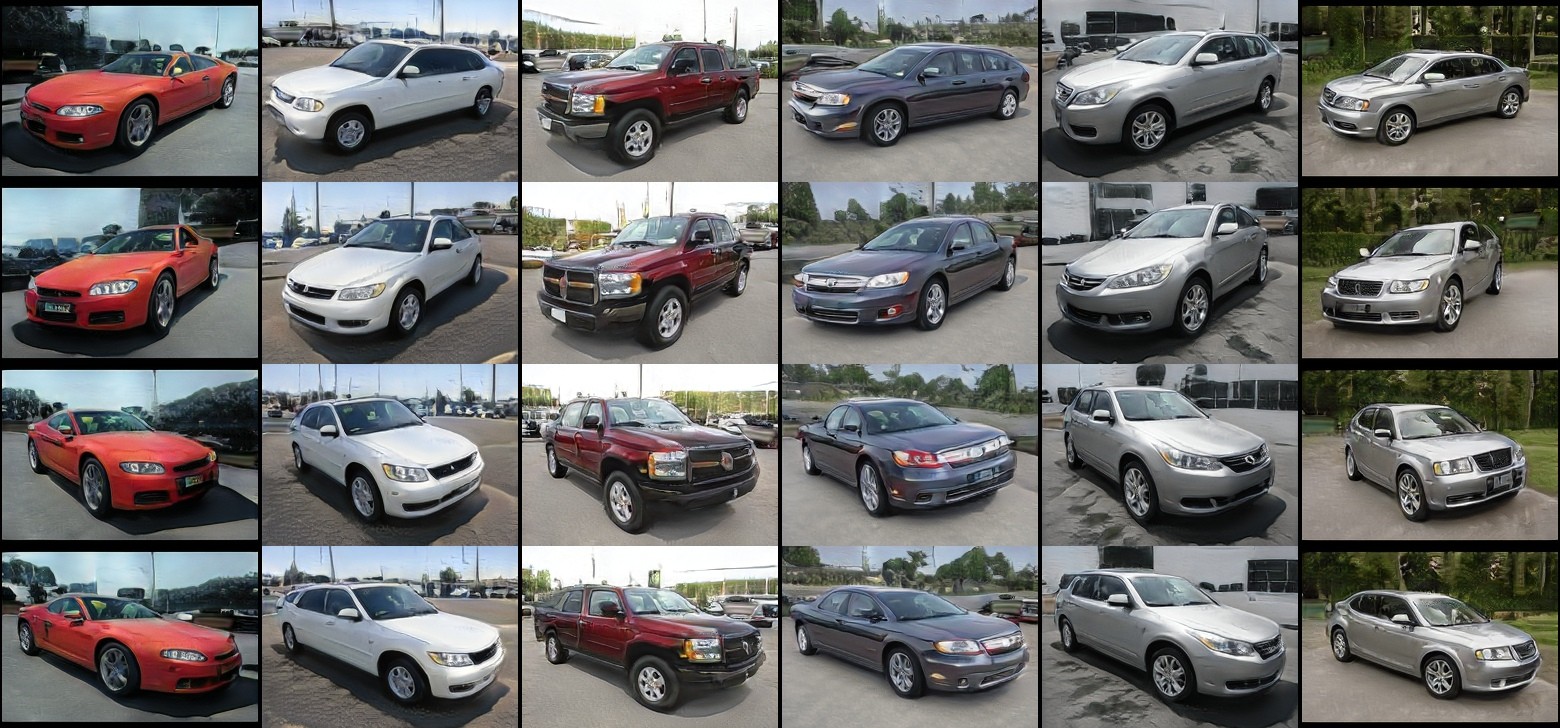}
    \end{tabular} \\
    $K=16$ &
    \begin{tabular}{c}
    \includegraphics[width=0.90\textwidth]{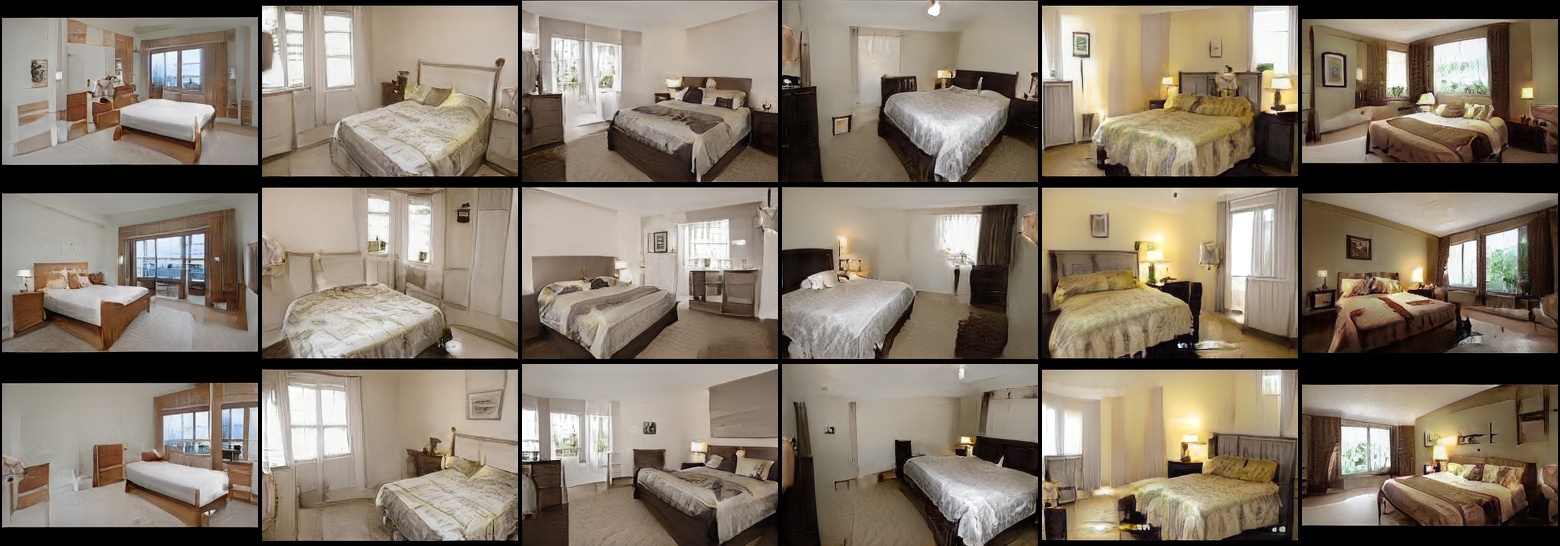}
    \end{tabular}
    \end{tabular}

    \caption{Disentanglement of mode and style:
    Samples generated by \UName{s} trained on four different datasets: handwritten letters `A' and `B' ($K=2$),
    Cats \& Dogs ($K = 2$), Cars ($K = 4$), and Bedrooms ($K = 16$). 
    For each set of samples, each row corresponds to a different fixed mode ($w_R$ is set to a one-hot vector), while each column corresponds to a different fixed style vector $w$ (mapped from random vectors $z \in \mathcal{Z}$).
    Note the similarity in visual attributes (font thickness, colors, and textures) along each column, despite the changes in the mode (`A' changes to `B', dogs to cats, pose of the car, pose and position of the bed).
	}
    \label{fig:disent}
\end{figure*}
\end{center}

\subsection{Disentanglement of mode and ``style''}

The mode and the style in a heterogeneous dataset are often naturally disentangled. For example, consider a dataset consisting of cats and dogs (two different content modes), where some of the cats and the dogs may have similar looking fur patterns, colors, etc.
In the StyleGAN architecture, some degree of disentanglement between different factors of variation is achieved by the mapping network that deforms the simply distributed latent space $\mathcal{Z}$ into the intermediate latent space $\mathcal{W}$, which may have a more complicated shape (see Figure 6 or \cite{karras2019style}).
However, the use of a single root constant makes it difficult to disentangle between variations in style and variations in the mode, which often correspond to content variations. By replacing the single root constant of StyleGAN by a combination of multiple learned root constants, we effectively enable the network to better achieve such disentanglement.


This ability is demonstrated in Figure~\ref{fig:disent}. Each group of images there was generated by our \UName trained on a different dataset: from top to bottom, a dataset of the handwritten characters `A' and `B' (curated by ourselves), the DRIT Cats \& Dogs dataset \cite{lee2018diverse}, the Stanford Cars dataset \cite{KrauseStarkDengFei-Fei_3DRR2013}, and the LSUN Bedrooms dataset \cite{yu2015lsun}.
In each group, each row corresponds to a different fixed mode (by setting $w_R$ to a one-hot vector), while each column corresponds to a different fixed style vector $w$, produced by the mapping network from a random vector $z \in \mathcal{Z}$. It may be seen that each column indeed preserves most of the visual attributes. For example, the thickness of the stroke in the letters dataset, the colors and patterns of the pet's fur, the shape and color of the car, and the colors and lighting of the bedroom.

The meaning of the mode may differ for each dataset: in the letters and Cats/Dogs datasets, the mode coincides with the semantic class of the image (whether it is the letter `A' or `B', or the species of the pet). 
It should be emphasized that while, for these datasets, the learned modes correspond to the semantic image classes, this is not guaranteed to be the case, in general. In other words, the modes learned by the generator might reflect a different clustering of the data than the one that might be more readily perceived by a human observer.
For example, in more diverse datasets, such as the Cars or the Bedrooms, the modes coincide with other dominant characteristics of the content: the pose of a car, the position and the pose of the bed in the room. Regardless, note that the mode indeed remains fixed along each row, and changes along each column.

\begin{figure*}[ht]
\begin{tabular}{|cc|c|c|c|c|c|c|c|}
\hline
 & & \textbf{Glow} & \textbf{StyleGAN} & \textbf{Two StyleGANs} & \textbf{\SName} & \textbf{\UName} & \parbox{2cm}{\centering\textbf{\SName} \\ \textbf{Half Dataset}} & \parbox{2cm}{\centering\textbf{\UName} \\ \textbf{Half Dataset}} \\ \hline
\multicolumn{1}{|c|}{}                           &
FID & 45.63 & 36.15 & 36.03 & 32.09 & 33.13 & 32.79 &  34.5 \\ \cline{2-9} 
\multicolumn{1}{|c|}{\multirow{-2}{3mm}{\rotatebox[origin=c]{90}{\textbf{Letters}}}}  &
\rotatebox[origin=c]{90}{Samples} & 

\begin{tabular}{c}
\includegraphics[width=0.04\textwidth]{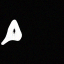} \includegraphics[width=0.04\textwidth]{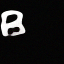} \\ \includegraphics[width=0.04\textwidth]{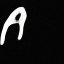} 
\includegraphics[width=0.04\textwidth]{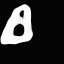} \\ \includegraphics[width=0.04\textwidth]{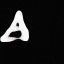} \includegraphics[width=0.04\textwidth]{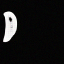}
\end{tabular}

&

\begin{tabular}{c}
\includegraphics[width=0.04\textwidth]{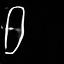} \includegraphics[width=0.04\textwidth]{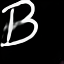} \\ \includegraphics[width=0.04\textwidth]{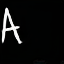} 
\includegraphics[width=0.04\textwidth]{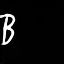} \\ \includegraphics[width=0.04\textwidth]{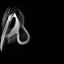}  \includegraphics[width=0.04\textwidth]{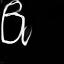}
\end{tabular}

& 

\begin{tabular}{c}
\includegraphics[width=0.04\textwidth]{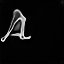} \includegraphics[width=0.04\textwidth]{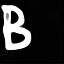} \\ \includegraphics[width=0.04\textwidth]{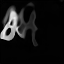} 
\includegraphics[width=0.04\textwidth]{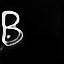} \\ \includegraphics[width=0.04\textwidth]{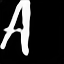} \includegraphics[width=0.04\textwidth]{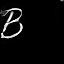}
\end{tabular}

& 

\begin{tabular}{c}
\includegraphics[width=0.04\textwidth]{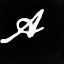} \includegraphics[width=0.04\textwidth]{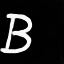} \\ \includegraphics[width=0.04\textwidth]{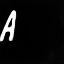} 
\includegraphics[width=0.04\textwidth]{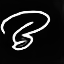} \\ \includegraphics[width=0.04\textwidth]{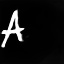} \includegraphics[width=0.04\textwidth]{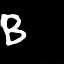}
\end{tabular} 

&

\begin{tabular}{c}
\includegraphics[width=0.04\textwidth]{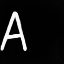} \includegraphics[width=0.04\textwidth]{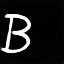} \\ \includegraphics[width=0.04\textwidth]{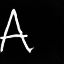} 
\includegraphics[width=0.04\textwidth]{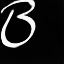} \\ \includegraphics[width=0.04\textwidth]{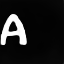} \includegraphics[width=0.04\textwidth]{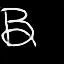}
\end{tabular} 

&

\begin{tabular}{c}
\includegraphics[width=0.04\textwidth]{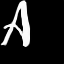}
\includegraphics[width=0.04\textwidth]{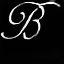} \\
\includegraphics[width=0.04\textwidth]{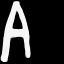}
\includegraphics[width=0.04\textwidth]{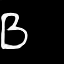} \\
\includegraphics[width=0.04\textwidth]{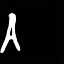}
\includegraphics[width=0.04\textwidth]{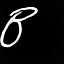} \\
\end{tabular} 

&

\begin{tabular}{c}
\includegraphics[width=0.04\textwidth]{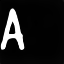}
\includegraphics[width=0.04\textwidth]{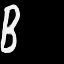} \\
\includegraphics[width=0.04\textwidth]{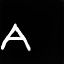}
\includegraphics[width=0.04\textwidth]{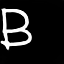} \\
\includegraphics[width=0.04\textwidth]{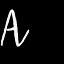}
\includegraphics[width=0.04\textwidth]{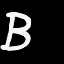}
\end{tabular} 

\\ \hline
\multicolumn{1}{|l|}{}                           &
FID & 123.75 & 126.90 & 120.44 & 107.94 & 117.25 & 109.11 & 119.13 \\ \cline{2-9} 
\multicolumn{1}{|c|}{\multirow{-3}{3mm}{\rotatebox[origin=c]{90}{\textbf{Dogs \& Cats}}}} &
\rotatebox[origin=c]{90}{Samples}

&

\begin{tabular}{c}
	\includegraphics[width=0.04\textwidth]{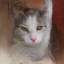} \includegraphics[width=0.04\textwidth]{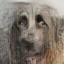} \\ \includegraphics[width=0.04\textwidth]{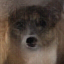} 
	\includegraphics[width=0.04\textwidth]{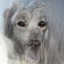} \\ \includegraphics[width=0.04\textwidth]{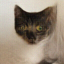} \includegraphics[width=0.04\textwidth]{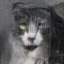}
\end{tabular}

& 

\begin{tabular}{c}
\includegraphics[width=0.04\textwidth]{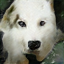} \includegraphics[width=0.04\textwidth]{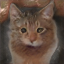} \\ \includegraphics[width=0.04\textwidth]{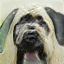} 
\includegraphics[width=0.04\textwidth]{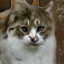} \\ \includegraphics[width=0.04\textwidth]{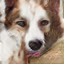} \includegraphics[width=0.04\textwidth]{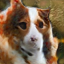}
\end{tabular} 

&

\begin{tabular}{c}
\includegraphics[width=0.04\textwidth]{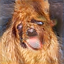} \includegraphics[width=0.04\textwidth]{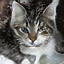} \\ \includegraphics[width=0.04\textwidth]{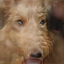} 
\includegraphics[width=0.04\textwidth]{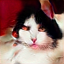} \\ \includegraphics[width=0.04\textwidth]{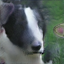} \includegraphics[width=0.04\textwidth]{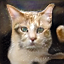}
\end{tabular} 

&

\begin{tabular}{c}
\includegraphics[width=0.04\textwidth]{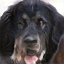} \includegraphics[width=0.04\textwidth]{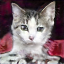} \\ \includegraphics[width=0.04\textwidth]{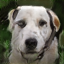} 
\includegraphics[width=0.04\textwidth]{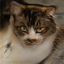} \\ \includegraphics[width=0.04\textwidth]{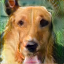} \includegraphics[width=0.04\textwidth]{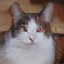}
\end{tabular} 

&

\begin{tabular}{c}
\includegraphics[width=0.04\textwidth]{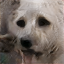} \includegraphics[width=0.04\textwidth]{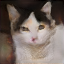} \\ \includegraphics[width=0.04\textwidth]{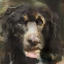}
\includegraphics[width=0.04\textwidth]{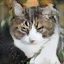} \\ \includegraphics[width=0.04\textwidth]{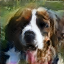} \includegraphics[width=0.04\textwidth]{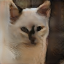}
\end{tabular} 

&

\begin{tabular}{c}
\includegraphics[width=0.04\textwidth]{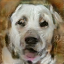}
\includegraphics[width=0.04\textwidth]{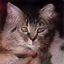} \\
\includegraphics[width=0.04\textwidth]{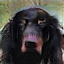}
\includegraphics[width=0.04\textwidth]{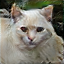} \\
\includegraphics[width=0.04\textwidth]{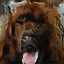}
\includegraphics[width=0.04\textwidth]{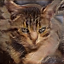} \\
\end{tabular} 

&

\begin{tabular}{c}
\includegraphics[width=0.04\textwidth]{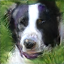}
\includegraphics[width=0.04\textwidth]{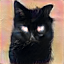} \\
\includegraphics[width=0.04\textwidth]{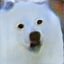}
\includegraphics[width=0.04\textwidth]{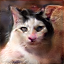} \\
\includegraphics[width=0.04\textwidth]{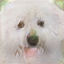}
\includegraphics[width=0.04\textwidth]{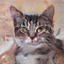} \\
\end{tabular} 

 \\ \hline
\end{tabular}
\caption{Performance of different generative models on two bi-modal datasets. From left to right, the models are: Glow, StyleGAN, two StyleGANs (each trained on part of the dataset containing a single mode), our supervised \SName, our unsupervised \UName, \rev{as well as \SName and \UName trained on only half of the training set}. The dataset in the top row consists of two letters `A' and `B', handwritten in a variety of styles. The dataset in the bottom row consists of portraits of cats and dogs (from DRIT \cite{lee2018diverse}). \rev{For the \UName or \SName models, we set the number of modes to $K=2$, reflecting the natural modalities exhibited by the letters and DRIT datasets.}
} 
\label{tab:FID}
\end{figure*}

\section{Experiments}
\label{sec:results}

We implemented the \OurName architectures discussed in the previous sections in PyTorch, \rev{by extending a publicly available implementation of StyleGAN \cite{StyleGANPyTorch}}.
It should be noted that the original StyleGAN of Karras \etal~\shortcite{karras2019style} is a special case of our \UName architecture, with $K = 1$.

\rev{The weights of the mapping network for $w_R$ are shared with the mapping network for $w$, except the penultimate fully connected layer, which is followed by a softmax operation. We found that separating the two mapping networks results in negligible visual benefit. We also found that feeding two separate $z$ vectors, one to each mapping network, results in decreased visual quality. 

Due to the weight sharing, the added parameters in \UName amount to the $512\times K$ parameters of the penultimate fully connected layer, which maps a vector of dimension 512 down to $K$. The added root constants (on top of the already existing single root) account for $4\times4\times512\times(K-1)$ additional parameters. The increase in training time as a function of $K$ is negligible.
}

For the StyleGAN and \OurName models, all of the results reported below were obtained by training for 200,000 iterations, in a progressively growing manner \cite{karras2017progressive}. Glow models were trained for 200,000 iterations. \rev{All of the models we compare to were trained using the hyper-parameters which were proposed by the original authors.} Note that we trained our models with 200,000 iterations due to limited computational resources.
\rev{Specifically, we used either GeForce GTX 1080 Ti or RTX 2080 Ti, with which it took {\raise.17ex\hbox{$\scriptstyle\mathtt{\sim}$}}2 days to train each model}.
Training for 2,000,000 iterations, as suggested by Karras \etal~\shortcite{karras2019style}, may further improve results.

For our quantitative comparisons between different generative models, we make use of the widely accepted FID metric \cite{heusel2017gans}. Each FID score is calculated on two sets (real and fake) of 5000 images.
Throughout this section, we refer to the (empirical) number of modes in our multi-modal datasets as $N$ (not to be confused with $K$, which denotes the number of constants in our models).
For all experiments on images of handwritten characters, the output of all generators, and the input of all discriminators, was changed to grayscale images of $64 \times 64$ pixels.
In all of our other experiments, the images/generators all use the resolution of $128 \times 128$, unless stated otherwise. Again, this resolution was chosen due to our limited computational resources, but it is not an inherent limitation.

We found the truncation trick, which was used by Brock \etal~ \shortcite{brock2019biggan} and Karras \etal~ \shortcite{karras2019style}, helpful for improving quality. Thus, all of the images shown hereinafter, from both our models and from the models we compare to, are produced using truncation.

\setlength{\tabcolsep}{1pt}
\begin{figure*}[]
	\centering
	\setlength{\tabcolsep}{2pt}
	\renewcommand{\arraystretch}{1.2}
	\begin{tabular}{ccc}
		\begin{tabular}{c}
			\includegraphics[height=2.2cm]{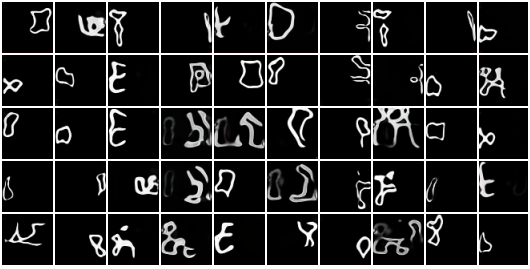} 
		\end{tabular} & 
		\Large
		\begin{tabular}{c|c|c|c|c|c|c|c}
			w & 8 & g & y & N & H & m & F \\ \hline
			B & X & M & 3 & g & 0 & 4 & B \\ \hline
			A & v & s & 4 & L & c & Q & Y \\ \hline
			0 & D & p & o & v & p & A & Y \\
		\end{tabular}
		\begin{tabular}{c}
			\includegraphics[height=2.2cm]{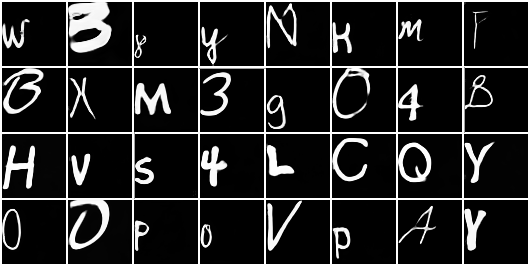} 
		\end{tabular} & 
		\begin{tabular}{c}
			\includegraphics[height=2.2cm]{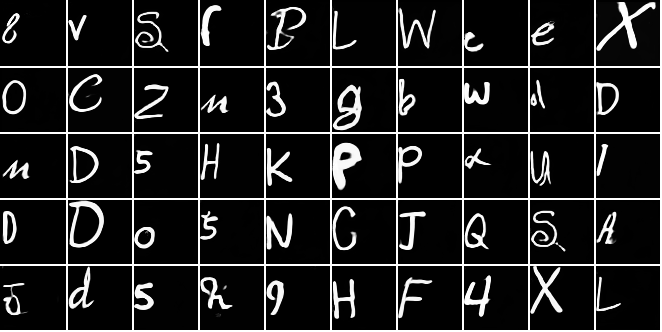} 
		\end{tabular} \\
		(a) StyleGAN, FID=61.90 & (b) mode inputs and results for \SName, FID=24.50 & (c) \UName, FID=41.35
	\end{tabular}
	\caption{Results of StyleGAN, \SName ($K = 62$), and \UName ($K = 62$), on a dataset of $62$ characters.
		(a) Samples generated by StyleGAN are clearly implausible. (b) the conditional mode inputs $w_M$ to \SName and the corresponding results. Adding multiple roots and supervision in \SName vastly improves the results. Note that $w_M$ properly controls the desired output mode. (c) The results of \UName are clearly better than StyleGAN, and only slightly degraded compared to \SName, due to the removal of the supervision. We also report the FID scores of each model, showing the benefit of multiple roots.}
	\vspace{-1em}
	\label{fig:StyleGANvsMultiConstGAN}
\end{figure*}

\subsection{Generation Quality}

We begin by comparing the visual and the quantitative quality of the results generated by StyleGAN \cite{karras2019style} and Glow \cite{kingma2018glow} to those generated by \SName and \UName, when trained on bi-modal datasets containing images from two classes: our own dataset of handwritten letters `A' and `B', and the Cats \& Dogs dataset from DRIT~\cite{DRIT_plus}. Since each of these datasets consists of two distinct classes, we also compare with two StyleGAN models, each trained on only the part of each dataset that consists of images from the same single class. \rev{Since training these two StyleGAN models involves splitting the training datasets, we also train \UName and \SName with only half of the dataset, to ensure a fair comparison.}
The results of this experiment are summarized in Figure~\ref{tab:FID}.

The visual results generated by Glow, appear to be blurrier than the training samples, and also compared to the results of the other methods in this comparison. Quantitatively, \rev{Glow's FID} scores are the worst for the A/B dataset, and second worst for the Cats/Dogs. It should also be noted that some of the images generated by Glow were completely black; we filtered these results out, before computing the FID scores.
Had these results been included, the resulting FID scores would have been much worse.

As for a single StyleGAN ($K = 1$), it may be seen that some of the images generated by it do not look like plausible handwritten letters `A' or `B'. This is also true for some of the Cats/Dogs results, 
where some of the generated images look like a mixture between a dog and a cat (see for example the bottom image in the fourth column from the left).
Quantitatively, the single StyleGAN results in the highest (worst) FID score, among the GAN-based methods, for both datasets.
When training two StyleGAN models on each dataset (split accordingly), the reduction in the FID score is quite small. We hypothesize that the reason for this is that the two separate models have more degrees of freedom, but they fail to leverage the style commonalities present across the entire multi-modal input distribution. An additional contributing factor is that each of the two StyleGAN models in this experiment was trained on only half the number of training samples.

Compared to the above alternatives, both our multi-modal architectures (\SName and \UName) exhibit better performance. The generated handwritten letters appear much more clear and plausible, while our Cats/Dogs results are more consistent and realistic. The FID scores for both our architectures are lower (better) as well. Naturally, the supervised variant, \SName, results in better FID scores than the unsupervised \UName. We attribute the improved performance of \UName compared to the unimodal StyleGANs to the fact that it is able to benefit from learning the joint visual attributes of both classes in each dataset, but without being forced to synthesize both classes from a single learned constant.
\rev{It may be seen that halving the training datasets of \SName and \UName slightly degrades the FID score. However, our supervised variant \SName persists to outperform all of the unsupervised variants. Additionally, \UName with a halved dataset still outperforms the single StyleGAN, as well as the two separate StyleGANs.}

We have also experimented with a dataset consisting of a large number of distinct classes.
Figure~\ref{fig:StyleGANvsMultiConstGAN} compares the results of StyleGAN to those generated by \SName and \UName, where all three architectures are trained on a dataset of consisting of 62 characters, handwritten in 216 different styles, with a total of ~14,000 images.
This dataset was manually curated by ourselves.

Figure~\ref{fig:StyleGANvsMultiConstGAN}(a) clearly demonstrates that the results of StyleGAN trained on a dataset with that many different modes are implausible, as almost none of the generated images resemble a character from the dataset. Figure~\ref{fig:StyleGANvsMultiConstGAN}(b) show the desired modes, as indicated by the mode vector $w_M$ and the corresponding results generated by \SName. The results are significantly better, and it is apparent that the desired mode is generated.
Figure~\ref{fig:StyleGANvsMultiConstGAN}(c) shows the results generated by \UName. While some of the results are slightly degraded compared to (b), they are still clearly more plausible than those of the unimodal StyleGAN. Additionally, we report the FID score obtained by each of the three models, showing that \UName improves upon StyleGAN and that, thanks to supervision, \SName outperforms both.

\begin{figure}[]
\large
\setlength{\tabcolsep}{6pt}
\renewcommand{\arraystretch}{1.2}

\begin{tabular}{c|c}
$K=1$ & $K=4$ \\
\hspace{-12px} 
\begin{tabular}{c}
\includegraphics[width=45px]{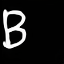} 
\end{tabular} 
\hspace{-12px}
&
\begin{tabular}{cccc}
\hspace{-12px} \includegraphics[width=45px]{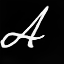} &
\hspace{-12px} \includegraphics[width=45px]{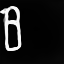} &
\hspace{-12px} \includegraphics[width=45px]{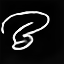} &
\hspace{-12px} \includegraphics[width=45px]{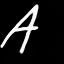} 
\end{tabular} \\
\end{tabular}

\begin{tabular}{c|c}
\begin{tabular}{c}
\hspace{-12px}
\includegraphics[width=45px]{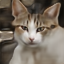} 
\end{tabular}
\hspace{-12px} 
&
\begin{tabular}{cccc}
\hspace{-12px} \includegraphics[width=45px]{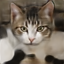} &
\hspace{-12px} \includegraphics[width=45px]{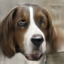} &
\hspace{-12px} \includegraphics[width=45px]{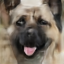} &
\hspace{-12px} \includegraphics[width=45px]{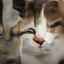} 
\end{tabular} \\
\end{tabular}

\begin{tabular}{c|c}
\begin{tabular}{c}
\hspace{-12px} \includegraphics[width=45px]{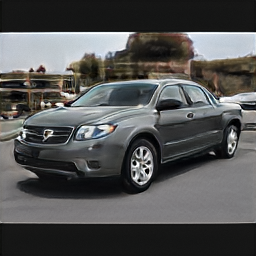}
\end{tabular}
\hspace{-12px}
&
\begin{tabular}{cccc}
\hspace{-12px} \includegraphics[width=45px]{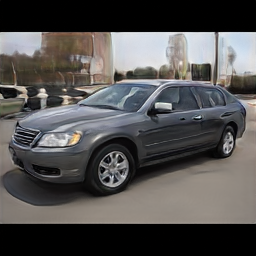} &
\hspace{-12px} \includegraphics[width=45px]{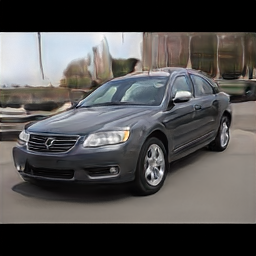} &
\hspace{-12px} \includegraphics[width=45px]{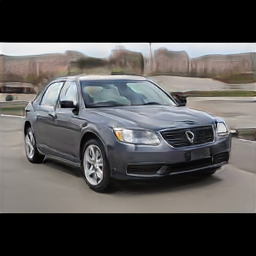} &
\hspace{-12px} \includegraphics[width=45px]{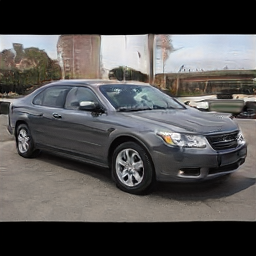} 
\end{tabular} \\
\end{tabular}

\begin{tabular}{c|c}
$K=1$ & $K=16$ \\
\begin{tabular}{c}
\hspace{-12px} \includegraphics[width=45px]{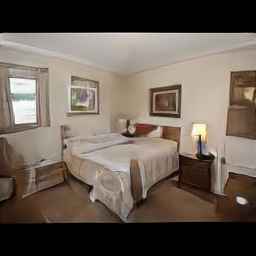}
\end{tabular}
\hspace{-12px}
&
\begin{tabular}{cccc}
\hspace{-12px} \includegraphics[width=45px]{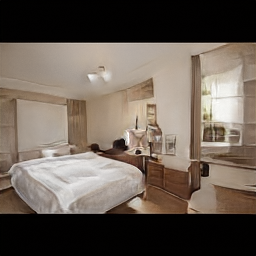} &
\hspace{-12px} \includegraphics[width=45px]{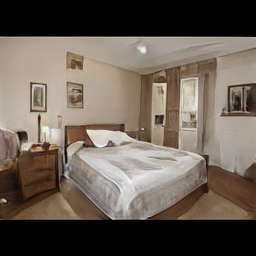} &
\hspace{-12px} \includegraphics[width=45px]{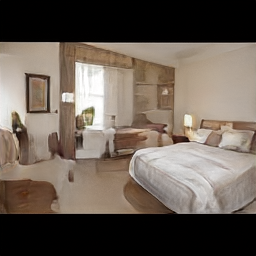} &
\hspace{-12px} \includegraphics[width=45px]{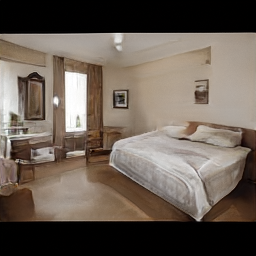} \\
\hspace{-12px} \includegraphics[width=45px]{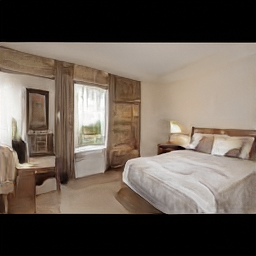} &
\hspace{-12px} \includegraphics[width=45px]{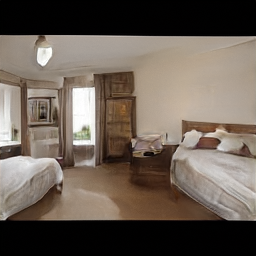} &
\hspace{-12px} \includegraphics[width=45px]{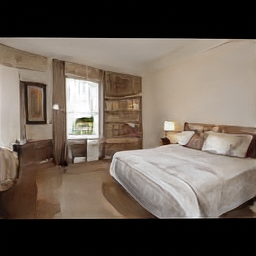} &
\hspace{-12px} \includegraphics[width=45px]{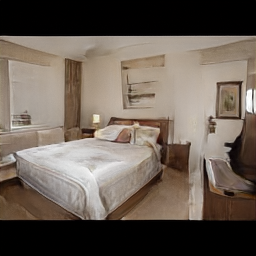} \\
\hspace{-12px} \includegraphics[width=45px]{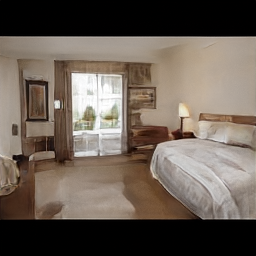} &
\hspace{-12px} \includegraphics[width=45px]{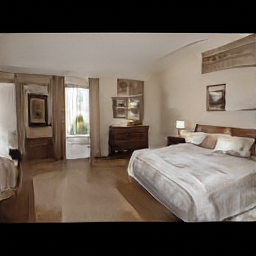} &
\hspace{-12px} \includegraphics[width=45px]{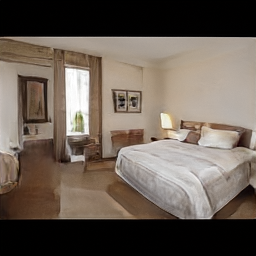} &
\hspace{-12px} \includegraphics[width=45px]{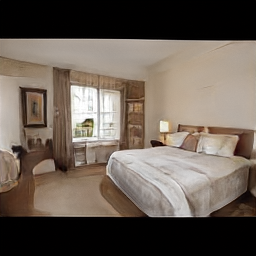} \\
\hspace{-12px} \includegraphics[width=45px]{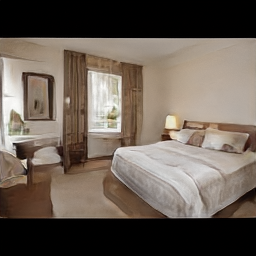} &
\hspace{-12px} \includegraphics[width=45px]{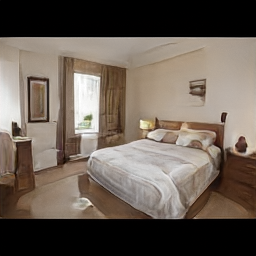} &
\hspace{-12px} \includegraphics[width=45px]{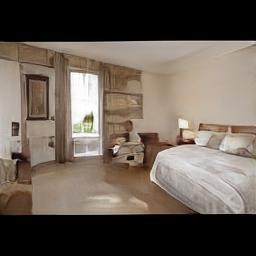} &
\hspace{-12px} \includegraphics[width=45px]{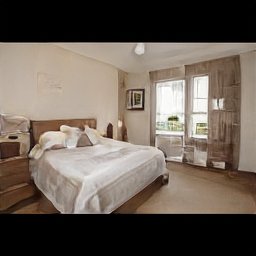} \\
\end{tabular}
\end{tabular}

\caption{Visualization of modes learned by \UName for different multi-modal datasets.
	In the top row, we show the modes for the handwritten letters, with $N=2$.
	In the next row, we visualize the modes for the bi-modal Cats \& Dogs dataset. As the number of modes increases, some of the modes might not correspond to a semantically valid image \rev{(for example the rightmost mode)}, but most correspond to typical images of cats or dogs. \rev{Nevertheless, the FIDs reported in Table \ref{tab:FIDvsK_DRIT} indicate an improvement as the number of modes is increased. A possible explanation for this behavior is that the additional modes enable some further refinement of results that could be obtained using a smaller number of modes.}
	For the Cars dataset, the modes correspond to different car poses, and for the Bedrooms dataset the modes mainly differ by pose and position of the bed in the image.
	\rev{It is noticeable that the 16 bedroom modes lack poses where the bed is oriented towards the viewer, as well as more zoomed-in views. We attribute this to the distribution of the LSUN dataset, in which such orientations and views are rare.}}
	\vspace{-1em}
\label{tab:VisConsts}
\end{figure}

\subsection{Learnt Modes}

At the basis of our approach is the assumption that the learned constants encode a template or a canonical representation of the modes present in the dataset from which we wish to sample and generate images.
In order to visually demonstrate this, we visualize the constants learned by \UName for several datasets. This is done by computing and applying an \emph{average style vector} $w_\textrm{avg} \in \mathcal{W}$, obtained by running 10,000 random latent vectors $z \in \mathcal{Z}$ through the mapping network of each generator, and averaging the results. Next, we disable the noise inputs to the synthesis network of the generator and apply the AdaIN parameters resulting from $w_\textrm{avg}$. Thus, the generator ``renders'' its learnt constant using an ``average'' style.
When we apply this method on \UName with $K > 1$, we force the generator to select one out of the $K$ learned constants by setting $w_R$ to a corresponding one-hot vector (note that the style vectors $w$ are generated independently of the mixing vectors $w_{R}$).

Figure \ref{tab:VisConsts} shows the resulting mode visualizations for our \UName with $K=1$, $K=4$ and $K=16$.
In the bi-modal datasets (letters `A' and `B', and Cats \& Dogs), it is apparent that StyleGAN (i.e., $K = 1$) has learned a root constant that represents one of the two classes present in the dataset (the letter `B' and a portrait of a cat).
This implies that in order to generate an image of the letter `A' or of a dog, StyleGAN must exert substantial effort in order to warp and texture the output properly, using only multi-scale AdaIN parameters.
Using larger values of $K$, \UName has learned very different modes, which better reflect the diversity of the dataset. In the case of the letters `A' and `B', and with $K=4$, the two learnt root constants yield images resembling the shapes of these two letters.

Similarly, when trained on the Cats \& Dogs dataset, the two constants correspond to images of a dog and a cat.
This effectively simplifies our generator's task of properly generating either one of the two classes.
We attribute the visual and quantitative improvements reported earlier to this simplification.
Recall that these root constants are learned in an unsupervised manner, in the course of the generator's adversarial training. In other words, our \UName is able to uncover two modes with which it spans the distribution of the dataset without any guidance.
Increasing the number of constants $K$ further, results in a more varied collection of ``templates'', and further improves the FID score, as will be discussed below.
Note, however, that the learned constants are not guaranteed to correspond to realistic, semantically meaningful images, as may be observed for the Cats \& Dogs dataset for $K = 4$.

Figure \ref{tab:VisConsts} also visualizes the learned constants for the datasets of Cars \cite{KrauseStarkDengFei-Fei_3DRR2013} and Bedrooms \cite{yu2015lsun}, which showcase images of the same semantic class, but with high intra-class visual diversity.
Here, we may observe that when multiple modes are learned by \UName, they correspond to different characteristic car poses for the Cars dataset, and for poses and positions of the bed in the bedroom images.

\begin{table}[]
	\large
	\setlength{\tabcolsep}{6pt}
	\renewcommand{\arraystretch}{1.2}
	\begin{tabular}{l|c|c|c|c}
		& $K=1$ & $K=2$   & $K=3$   & $K=4$   \\ \hline
		$N=2$ & 36.15 & \cellcolor[gray]{0.9} 33.13 & 32.56 & 31.62 \\ \hline
		$N=3$ & 36.80 & 34.41 & \cellcolor[gray]{0.9} 32.94 & 32.54 \\ \hline
		$N=4$ & 37.11 & 36.78 & 36.66 & \cellcolor[gray]{0.9} 34.98
	\end{tabular}
	\caption{The effect of the number of constants $K$ learned by \UName, as a function of the number of classes $N$ in the training set. For each combination of $N$ and $K$ we report the FID score.
		By examining the scores in each row, it is apparent that for a fixed number of classes, increasing $K$ improves the score. The largest improvement in the FID score is achieved when we move from $K < N$ to $N <= K$ (the gray cells on the diagonal). Additionally, by examining the columns we may notice that increasing the number of classes in the dataset for a fixed value of $K$ hurts the performance.}
	\label{tab:FIDvsK}
\end{table}

\subsection{Number of Modes}

In Table \ref{tab:FIDvsK}, we report the FID scores obtained by training our \UName with varying numbers of root constants ($K \in \{2,3,4\}$), i.e., learned modes, on datasets with varying numbers of classes ($N \in \{2,3,4\}$). Each of these datasets consists of images of $N$ different letters. It is each composed of 216 images of each letter type.

By examining the FID scores in each row of Table \ref{tab:FIDvsK}, it is apparent that for a fixed number of classes (modes) in the dataset, increasing the number of learned modes $K$ improves the results. This behavior is similar to that of clustering algorithms, where increasing the number of clusters typically results in a better fit of the data. 
It is interesting to note that the largest improvements occur when we move from a model where $K < N$ to one where $N <= K$. For example, for the case $N=3$ a drop of 1.47 in the FID score is observed when switching from $K = 2$ to $K = 3$, while increasing to $K = 4$ results in only a modest further reduction (0.4). The same behavior may be observed in the last row ($N = 4$).

By examining the columns of Table \ref{tab:FIDvsK} one may notice that increasing the number of classes in the dataset, while keeping the same number of learned modes $K$, results in reduced performance. We believe that this is caused by the increase in the generator's difficulty to cope with the added complexity and richness of the dataset. We stress that in this experiment, a dataset with a smaller value of $N$, is a subset of those with a larger $N$. For example, when increasing $N$ from two to three, we simply add another letter to the two already existing ones. This property is important, since it allows us to add new classes, without altering any property of the existing ones.

In Table \ref{tab:FIDvsK_DRIT}, we report the FID scores obtained by training our \UName with varying numbers of root constants ($K \in \{1,4,8,16\}$) on the Cats \& Dogs, Cars, and Bedrooms datasets. It is easily noticeable that, similarly to clustering methods, increasing the number of roots improves the FID score, due to the ability to more closely fit the data distribution observed in the training set. Note that the FID scores for these datasets are larger than those in Table \ref{tab:FIDvsK}, due to the increased visual complexity of the natural images.

\begin{table}[tb]
\large
\setlength{\tabcolsep}{4pt}
\renewcommand{\arraystretch}{1.2}
\begin{tabular}{l|c|c|c|c}
& $K=1$ & $K=4$   & $K=8$ & $K=16$   \\ \hline
Cats \& Dogs  & 126.90 & 109.42 & 106.44 & 103.49 \\ \hline
Cars          & 167.22 & 163.18 & 153.48 & 145.54 \\ \hline
Bedrooms      & 189.53 & 185.75 & 178.95 & 157.01 \\ \hline
\end{tabular}
\caption{The effect of increasing the number of constants $K$ learned by \UName, on three datasets of natural images. For each value of $K$ we report the FID score.
It is apparent that increasing $K$ improves the score, similarly to unsupervised clustering methods.}
\vspace{-1.5em}
\label{tab:FIDvsK_DRIT}
\end{table}

\section{Conclusions}

We have presented a novel architecture which targets diverse multi-modal distributions, without making any \emph{a priori} assumptions regarding the form of the distribution.
Our method has been shown to learn modes that may be used to span the target distribution observed in the training dataset, in an unsupervised fashion, requiring only the number of modes to be specified.

An important property of our multi-modal architecture is that it disentangles the mode from the other visual attributes. In that sense, it echoes well the natural disentanglement that usually exists between these properties in the training data. Thus, our architecture provides an additional degree of control over the generated samples, enabling to control the mode, which often corresponds to high-level content, independently from the remaining visual attributes, which may be thought of as style. 

We have shown that explicit representation of multiple modes aids the generation process and improves the quality of the generated distribution.
However, a current limitation of our approach is that the number of modes $K$ must be determined before the training. Thus, a natural direction for future work would be to extend our method  to also learn the optimal value of $K$.
This direction involves additional interesting questions, such as, a better understanding of the trade-offs involved in increasing $K$, \rev{as well as the impact of setting $K\ll N$ or $K\gg N$}.
Clearly, an overly large $K$ has a diminishing return, and might lead to overfitting.
\rev{Using an overly small $K$ most probably improves upon using $K=1$, since any increase enables more carefully fitting the underlying distribution.
Additional architectural optimizations should also be explored.}

The ability to control mode separately from style adds a new ``control axis'' to the style mixing controls available in the StyleGAN architecture. However, regardless of the number of modes $K$, in our experience they seem to be homing on a single dominant characteristic, such as animal species, or object pose.
It would be useful to be able to specify which characteristic should the modes capture, and even more useful to extend the mechanism to provide multiple control axes, rather than a single one.

\bibliographystyle{ACM-Reference-Format}
\bibliography{main}

\end{document}